\documentclass[11pt]{article}
\usepackage[margin=1in]{geometry}
\usepackage{amsmath}
\usepackage{booktabs}
\usepackage{tabularx}
\usepackage{graphicx}
\usepackage[hidelinks]{hyperref}
\usepackage{microtype}
\title{AZOWL AgentOps\\
\large Task-Scoped Governance for Deployed Agent Applications}
\author{Taeoh Hong\\BlueTeamK}
\date{19 November 2026 \quad\textperiodcentered\quad Manuscript draft}

\begin{document}
\maketitle

\begin{abstract}
AZOWL AgentOps provides shared governance for deploying agents and delegating work to them. Human and agent senders can exchange messages or explicitly delegate a business task. Accepted work has a stable identifier that links the sender's authority, the assigned runner, any required approval, and result evidence. The default enrolled customer PC profile is read-only. We compare AZOWL components with a configured LangGraph application in 28 synthetic runs across four paired pilots. Original AZOWL services checked work admission, recipient routing, and workflow prerequisites in isolated tests. The configured LangGraph application supplied business task and dispatch records in application code. Both configurations recorded zero business tasks in dialogue, rejected unauthorized agent handoffs, and issued zero writes when an approved delivery address changed. AZOWL's scope gate, exercised as a component, requested approval again on the same task. A product probe reused one registered provider across two workflows and checked prerequisite enforcement for ticket-task submission. The paired protected writes, including ticket creation, used mock providers. Of two deployed PC inventory reads with verified receipts, we traced the 0.6.57 run from submission through runner pickup and tool execution to the receipt, with matching local and central artifact hashes. A separate deployed GRC audit followed an operator-driven ticket lifecycle on sample records.
\end{abstract}

\section{Introduction}

An agent reviewing a report may ask a peer to interpret a finding or delegate delivery of the report. An interpretation is a conversational reply. Delivery is an assigned business objective that may require a decision about the exact report and destination. An operator should be able to trace that decision to the assigned runner and the observed result. Process compliance depends on the applicable rules and evidence of the actions taken~\cite{ly2015compliance}. Inventory collection and ticket creation present different business logic with similar needs for assignment, authority, and evidence.

Task-based authorization connects permissions to the progress of work~\cite{thomas1998tbac}. Existing studies test approval, refusal, restoration, and recovery from unacknowledged effects~\cite{deng2026auditable,sah2026undobench}. We study whether application developers can reuse governance services across agent applications and whether operators can use the resulting records to oversee deployed work. A task connects a work request to its assignment, authority, and evidence.

We present \textbf{AZOWL AgentOps}, a framework for building and operating governed agent applications. Developers specify workflow steps, dependencies, and registered execution providers. Across human-to-agent and agent-to-agent interactions, the platform separates an exchange of information from an explicit delegation of work. Accepted delegations enter the task lifecycle. The default enrolled PC profile runs in read-only prompt mode. Runners on enrolled hosts receive assigned tasks from the central service, with each protected action subject to the applicable authority. Central services expose runner status, task decisions, and result evidence. Section~\ref{sec:framework} describes the construction interfaces and interaction modes.

We evaluate AZOWL alongside a LangGraph application configured to meet the same business requirements. LangGraph supplies graph execution, persistence, and interrupts. The comparator uses application code for business task intake, dispatch, and scoped control. We ask four questions about the configured systems.

\begin{enumerate}
    \item Does each configuration preserve the boundary between conversation and delegated work for both human and agent senders, and which controls enforce it?
    \item Which controls can be reused to compose and deploy different governed applications, and what additional configuration or code does each application require?
    \item Does work reach the designated deployed runner, and can operators verify its assignment, applicable approval, and result evidence from the task identifier?
    \item Under changes in assignment or approval and under restart, does the application preserve task-scoped authority and resolve the earlier effect?
\end{enumerate}

This paper contributes a task-centered architecture that separates conversation, work admission, assignment ownership, and permission for a protected action. It presents four paired laboratory pilots and identifies which controls came from the platforms, the applications, or the shared apparatus. Separate deployment records show a native agent-to-agent reply, receipt and execution of bounded inventory tasks, approved report generation, and an operator-driven GRC ticket lifecycle on sample records. We use these records to distinguish observed component behavior from claims about deployed systems.

\section{Related Work}

\paragraph{Process compliance and work-based authorization.}
Ly et al. characterize compliance monitoring in terms of the rules and runtime observations available for a business process~\cite{ly2015compliance}. Thomas and Sandhu's task-based authorization controls associate permissions with the progressing state of a task~\cite{thomas1998tbac}. AZOWL applies these concerns to governance work with an LLM agent. The organizational control identifies the evidence the work should produce, and the task's authorization restricts the actions the assigned runner may perform.

\paragraph{Evaluation of agent frameworks.}
AIOS compares success rates, throughput, and latency with and without its runtime layer~\cite{mei2025aios}. AFlow compares generated workflows with manual and automated baselines on benchmark tasks and analyzes cost versus performance on HumanEval~\cite{zhang2025aflow}. LangGraph models workflows with shared state, nodes, and edges~\cite{langchain2026platform}. LangSmith Agent Server uses threads to retain state and runs to represent invocations~\cite{langchain2026runs}. Queue workers acquire leases on runs~\cite{langchain2026server}. The A2A endpoint uses task identifiers to refer to work within a conversation~\cite{langchain2026a2a}. LangSmith provides tracing and monitoring~\cite{langchain2026observability}. Our comparison counts tasks admitted to the business assignment and authorization lifecycle and checks how each configuration links authority, execution identity, and results.

\paragraph{Runtime authorization.}
AgentSpec evaluates a rule language for runtime enforcement across tool-using agents~\cite{wang2026agentspec}. FORGE uses action provenance and a reference monitor to enforce policy~\cite{palumbo2026forge}. Auditable Authorization evaluates approval, refusal, restoration, and scope revision~\cite{deng2026auditable}. Reconstruction-aware repair analysis follows how the approved representation becomes the action consumed by a protected service~\cite{zhu2026reconstruction}. Commit-time authorization evaluates which justification applies at commitment~\cite{santos2026temporary}. Our evaluation follows the related controls from task intake to evidence review.

\paragraph{Recovery and external effects.}
The $\tau$-bench evaluation tests multi-turn policy-guided tool use and checks the final database state against an annotated goal~\cite{yao2025taubench}. RIFL uses durable completion records and stable request identity to suppress duplicate effects under retry~\cite{lee2015rifl}. UndoBench separates ordinary competence from recovery and compares recovery methods after an unacknowledged commit, including in a LangGraph implementation~\cite{sah2026undobench}. The AZOWL stress test uses a related fault schedule and observes the approval and runner assignment in force for any recovery dispatch. LangGraph provides checkpoints, interrupts, and guidance on retry boundaries and idempotency~\cite{langgraph2026fault}. The configured reference application uses these primitives with explicit authorization and reconciliation logic.

\section{AZOWL Framework and Evaluation Boundary}
\label{sec:framework}

\begin{figure}[!htbp]
\centering
\includegraphics[width=\textwidth]{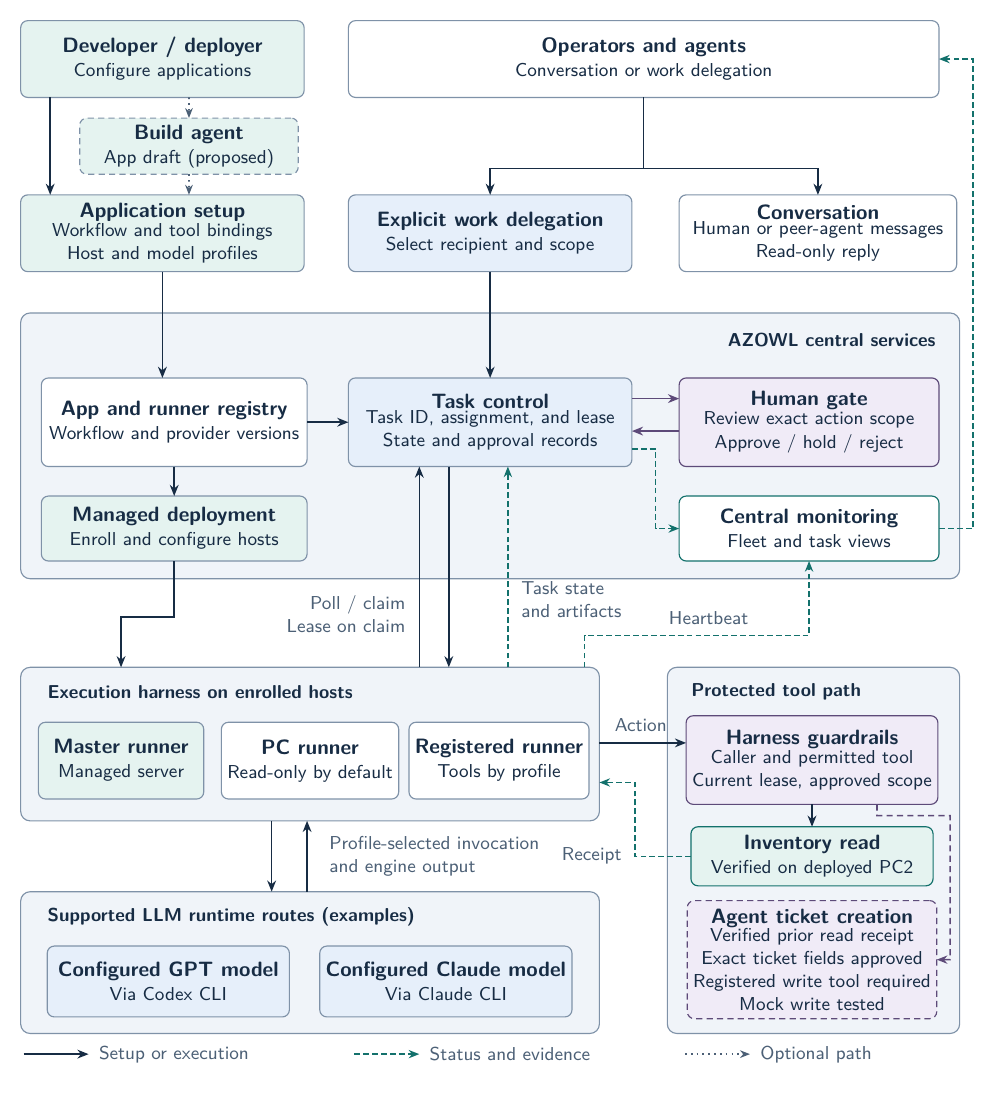}
\caption{Logical architecture of AZOWL. Profiles select model routes and tools. The inventory branch was verified on a deployed PC (Table~\ref{tab:deployed-read-trace}). The dashed agent ticket branch shows the design exercised with a mock write in the paired pilot. The separately audited GRC ticket lifecycle used the operator path.}
\label{fig:governance-framework}
\end{figure}

Figure~\ref{fig:governance-framework} shows how application setup and explicit work submission meet at the registry and task control service. Runners claim tasks, invoke their configured model runtime through the execution harness, and report state, artifacts, and heartbeats. The ticket design requires a verified predecessor receipt, authorized human approval of the exact ticket fields, and a registered write tool. It also requires checking the runner's current assignment lease and the approved scope before dispatch. The paired pilot tested ticket writes through a laboratory connector and a mock provider.

\subsection{Application composition and deployment}
\label{sec:application-composition}

In the reviewed shared-workflow implementation, a developer declares steps with objectives, inputs, required tools and outputs, and acyclic dependencies. Each step can be bound to a registered execution provider that identifies an agent or runner and its permitted tools. Dispatch checks the workflow, step binding, and provider revisions. A step with prerequisites requires verified predecessor runs. A connected host runs the assigned task and returns execution evidence through the shared runtime.

The application determines its work and inputs. The common services supply registration, task identity, assignment, and central review. Policy and tool support govern the actions available on each path. The shared-workflow definition supports acyclic dependencies and registered providers and adapters. The live example below used an inventory-read adapter. It verified one workflow step executed by a registered PC runner.

\paragraph{Task delivery to a deployed agent.}
The workflow binding names a registered execution provider. The central service queues the task with the target runner and applicable routing constraints. For the native PC path, the runner on the enrolled host polls for eligible work and claims the task. Central task control records the claim and issues a lease to that runner. The runner starts the assigned agent through its execution harness and reports state and results against the same task identifier.

\paragraph{Proposed Build agent.}
A Build agent could draft workflow steps and dependencies from a developer's description and propose tool and execution-provider bindings. The developer would review the draft and submit the application definition through the existing registration and deployment path. At execution, the task and its assignment would govern access to the requested tools. The proposed authoring assistance is outside the scope of the current evaluation.

\paragraph{Master runner in managed deployment.}
When an organization requests a managed server environment and its deployment is approved, the deployment manifest specifies a tenant-specific Master runner and profile by default. The managed workspace submits cloud tasks to the runner in its deployment binding. A shared-workflow step can be submitted directly to its registered runner through its execution-provider binding. Central task control records assignments, leases, and on-task decisions for the applicable work. A deployment check confirmed service readiness for the Master runner and a separate mission controller on a managed Linux host.

\paragraph{LLM runtime integration.}
The workflow's execution-provider binding identifies the runner. Its profile supplies the engine, model identifier, and reasoning setting where supported. The reviewed native executor invokes the corresponding Codex or Claude CLI and receives its execution output. Tool authority remains attached to the task and the integration that executes the action. Configuration records identify the requested settings. Invocation evidence is needed to establish which route ran and, where reported by the provider, which model version actually served it.

\subsection{Example governance workflow}

\begin{enumerate}
    \item \textbf{Discuss the review.} Ask what evidence is needed for the asset inventory control. Use the conversation to clarify the relevant organization, evidence sources, and expected output.
    \item \textbf{Submit and monitor the work.} Request an inventory read for the specified organization and ask for the result and its source record. On acceptance, follow the assignment and progress by task ID. The fleet view shows when the assigned runner last reported.
    \item \textbf{Respond and review.} If a runner needs more information, supply it on the task. If policy requires an approval, an authorized reviewer records a decision through the applicable approval path. Inspect the result and the evidence returned for the task.
\end{enumerate}

In the bounded read qualification, a designated runner retrieved eight assets for the authorized organization. The backend receipt matched the runner evidence for the same task. A completed collection gives the reviewer evidence to assess the control against the applicable criteria.

\subsection{Conversation and work delegation}

The application interaction model includes both conversation and explicit work delegation. An agent can ask a peer what an inventory result implies and receive a reply within a conversation. An authorized agent can commission that peer to inspect an asset and return evidence through the work admission path. Accepted work creates a governed task with a stable identifier, an assignment, and any applicable approval. The execution harness checks authority on protected tool paths.

The native agent-to-agent (A2A) message path carries a read-only request and reply within authorized sender and recipient scopes. The reviewed portal Master service also exposes distinct message and work-handoff operations. The handoff service checks the delegated caller's work-creation permission and binds accepted work to the selected recipient. Earlier portal message transport used a read-only execution record to carry a reply. The direct message path uses the A2A channel. We count a governed task when a business objective enters the assignment and authorization lifecycle. The executable-step schema is a separate part of the shared-workflow service.

\subsection{The task as the unit of execution authority}

A workflow organizes related work. An individual task gives one assignment a stable identity, an ownership record, and a state history. An assignment lease limits which runner currently owns that task. Authorization at the execution boundary additionally evaluates the authenticated caller and the requested tool and action within that task. Approval, when required by policy, supplies a human decision about a particular action scope. An operation identifier describes one intended business effect within the task and remains stable across a retry. Table~\ref{tab:authority} distinguishes these records by what each establishes.

\paragraph{Read-only defaults and authority to act.}
The reviewed customer PC provisioning path starts the runner in \texttt{prompt\_readonly} with an empty tool allowlist and no action grants. Standard SaaS work submission disables operator execution tools, including on formally assigned PC tasks. Work creation is checked against the sender's permission. A claim records the recipient's ownership of that task. Protected tool execution requires a profile and integration that permit the requested tool, current task authority, and any required scoped decision.

\paragraph{Execution harness and guardrails.}
Workspace access uses role-based access control (RBAC). Registered runner roles also define tool allowlists. On the protected inventory route, the execution harness carries task context to the tool interface, where the integration checks the authenticated caller, permitted tool, task scope, and current assignment lease before dispatch. Enforcement depends on the path. Selected direct commands have a pre-execution approval check. The inspected agent-shell monitor can flag activity after execution. The evaluation records which checks block dispatch and which produce audit observations.

\paragraph{Human decisions.}
A human gate holds a protected action for an authorized decision where required. An approval makes the action eligible to proceed subject to the current task, scope, and ownership checks. A hold or rejection keeps it stopped. The task service can record questions and answers, which the runner consumes through the applicable resume path.

\begin{table}[htbp]
\small
\centering
\begin{tabularx}{\textwidth}{@{}lX@{}}
\toprule
Record & What it identifies or establishes \\
\midrule
Task and organization & The task ID, organization or workspace, task revision, and lifecycle status. \\
Work creation permission & The authenticated sender's authority to submit a work delegation. \\
Runner identity and lease & The runner, lease ID, and expiry used to establish current assignment ownership. \\
Execution profile & The tool allowlist and configured grants for the receiving runner. \\
Question and answer & A question ID, its open or answered state, and the input or decision recorded in the response. \\
Action scope and approval & The authorized action and the scope and current validity of any required approval. \\
Operation and effect receipt & The intended business effect and the available evidence of its commitment. \\
\bottomrule
\end{tabularx}
\caption{Records the evaluation uses to distinguish task progress, assignment, input, authority, and external effects. Availability and checks depend on the execution path.}
\label{tab:authority}
\end{table}

\subsection{Task lifecycle}
\label{sec:task-lifecycle}

Table~\ref{tab:lifecycle} shows selected status values from the pinned central service and native runner. The standard intake creates a \texttt{pending} task. A successful claim records \texttt{claimed} and issues a lease. A runner can record execution, a wait, or a terminal outcome. In the central service, releasing a task can return it to \texttt{pending}. For a resumable prompt assignment, the native runner can use an accepted answer to make the local assignment \texttt{prompt\_ready} again. The work path determines which recorded values and transitions apply.

\begin{table}[!htbp]
\small
\centering
\begin{tabularx}{\textwidth}{@{}p{0.21\textwidth}p{0.30\textwidth}X@{}}
\toprule
Stage & Selected values & Meaning in the work path \\
\midrule
Intake and readiness & \texttt{pending}, \texttt{prompt\_ready} & Queued for dispatch or ready for the prompt executor. \\
Assignment & \texttt{claimed} & The central service has recorded a runner claim and issued a lease. \\
Execution & \texttt{running} & The runner reports execution in progress. \\
Waiting or blocked & \texttt{needs\_input}, \texttt{blocked} & The task is waiting for input or has a recorded condition preventing progress. \\
Terminal & \texttt{completed}, \texttt{failed}, \texttt{cancelled}, \texttt{superseded} & The central task store recognizes completion, failure, cancellation, or replacement. \\
\bottomrule
\end{tabularx}
\caption{Selected AZOWL work-task status values. Stage names group values for exposition.}
\label{tab:lifecycle}
\end{table}

The central answer API records the change in question state from \texttt{open} to \texttt{answered}. Resuming the assignment is a separate runner action. A protected action also requires the applicable authorization when it executes. In the SaaS interface, a task may still display \texttt{running} while the service waits for the result artifact from a task reported as \texttt{completed} by the native runtime. The evaluation retains the source of each status and checks completion against independent business evidence. After an interrupted request, the provider may already have committed an effect while the worker is still waiting for confirmation.

\subsection{Monitoring deployed agents}
\label{sec:deployment-monitoring}

In a managed deployment, host enrollment and the registered profile give operators a reference for the deployed agent, its runner, and the host on which it runs. The runner sends heartbeats to the central runtime service. The fleet view groups reported runners by host and shows runner and team identifiers, last-seen time, and freshness. The operator can check whether a runner is still reporting while following a separate task by its ID.

The task view draws on task and run records for assignments, progress, outstanding questions, and returned results. Authorized operators can also review recorded policy and operator decisions through the governance ledger view. Together, these views let an operator investigate why work is waiting, who is assigned, and which evidence is available for the result. A heartbeat establishes recent reporting. Evaluating the action or the control requires the corresponding authorization and evidence records, including a service receipt where available.

\paragraph{Evaluated AZOWL paths.}
In the bounded system-read path, the central tool endpoint authenticates the runner and checks its registered role, tool permission, task, and current lease. The backend verifies the endorsed request against the organizational binding and sealed run command and rechecks the lease. In the deployed compiled-plan report path, the approval question binds the plan revision and selected input artifact digests. The native runner uses the recorded decision to resume the report task. The paired write pilots connect original admission or approval components to isolated business adapters and retain the provider's effect ledger.

\paragraph{LangGraph reference application.}
The reference application combines LangGraph interrupts and a persistent SQLite checkpointer with application code for business admission, scoped approval, artifact-digest checks, and an operation journal. The A2A pilot also runs separate sender and recipient graphs. The paired write pilots use a common isolated approval route and record assignment ownership at the provider boundary. The configured reference uses a common rule for reconciliation lookup under its test access policy. The work-admission identity and execution surfaces used in each configuration are documented in the method and results.

\paragraph{Independent observation.}
Experiments use synthetic organizations, accounts, inputs, and loopback business services. The evaluated application sees the same provider API as its matched comparator. The independent grader reads the expected-effects manifest, committed-effect ledger, and final state. Expected answers and privileged ledger inspection are available to the grader and supervisor. The study assumes a trusted supervisor and grader and a fail-stop process failure on a working host. Provider capability, evidence visibility, and the evaluated authorization boundary are recorded for every batch.

\section{Evaluation Protocol}

We ran four source-pinned paired pilots on PC2. The human intake pilot comprised six runs, the agent-to-agent pilot eight, the workflow reuse pilot six, and the response-loss pilot eight. Each condition in each batch used one run per configuration. Matching held the synthetic business inputs and expected effects fixed. The write paths used the same laboratory approval and provider interfaces within each pair. Product interfaces and substitutions are recorded for each path. Each completed batch includes the scenario inputs, version or source fingerprints, an independently checked task and provider trace, and a run-level verdict. We interpret the 28 runs descriptively and retain the separate batch denominators.

\subsection{Cross-application construction and reuse}
\label{sec:construction-reuse}

The paired reuse pilot evaluated a standalone inventory read, a ticket that depended on the read receipt and exact approval, and a ticket whose assignee changed after approval. The observer checked the predecessor receipt against the task history, the ticket against its approved fields, and the final mock service state against the expected effect. An isolated AZOWL product probe separately exercised registration of a provider, binding the same version to both workflow definitions, and submission of the dependent step with a missing, unverified, or verified prerequisite. The probe used a fixture predecessor result and captured the product's dispatch request. The paired ticket write used a laboratory connector.

We attribute each control to the code that evaluated it. The isolated product routes ran original admission, routing, and prerequisite checks over fixture inputs. Native approval functions invoked through a laboratory bridge provide evidence about those components. The supervisor supplied the synthetic task inputs, captured dispatch transports, arranged the linked laboratory assignment, and connected protected actions to mock providers. The local LangGraph app used graph and checkpoint interfaces with application code for work admission, scope checks, and journaling. Table~\ref{tab:control-placement} separates these contributions.

\subsection{Conversation, work intake, and operator oversight}
\label{sec:workflow-utility}

\paragraph{Matched intake boundary.}
The human intake pilot used the same report request for discussion, explicit work followed by approval, and a recipient change after approval. The AZOWL path authenticated a synthetic account through the product routes. The LangGraph app used the corresponding caller supplied by the laboratory. The observer recorded the selected entry mode, work-task and dispatch records, and any provider request. The expected outcomes were zero work tasks and writes in discussion, one accepted task and one approved effect for the authorized submission, and zero writes for changed scope. The approved-work check also inspected the provider during approval wait. Provider ingress was counted before deduplication.

\paragraph{Agent-to-agent interaction control.}
The agent-to-agent pilot held the report request and sender and recipient roles fixed. The experiment selected dialogue, authorized handoff, handoff without work-creation authority, or recipient change after approval. The AZOWL path exercised the signed Master callback, the direct device messaging route with a fixture reply, and the original work admission route with a captured dispatch. The LangGraph app invoked separate sender and receiver graphs, with a receiver process and durable application task records. The observer traced the message and reply, accepted work and intended recipient, and any approval, request, or effect in the laboratory path. Business task counts reflect admission of a work objective to the assignment and authorization lifecycle. Conversation and transport records were inspected separately.

\paragraph{Assignment and effect evidence.}
The completed intake pilots recorded the AZOWL product handoff as read-only. The supervisor carried that identifier into the isolated central service for a separate protected write test and arranged a runner claim there. The isolated approval route validated the test user's decision. The separately connected gate and adapter then determined whether to call the mock provider. The observer graded completion against the expected effect and linked the applicable decision, current assignment, attempted action, and provider receipt.

\subsection{Authorization and operation identity}

The evaluation manifest records the authenticated approval where applicable, workspace, task, subject, approved tool, target and arguments, artifact digest, policy revision, and logical operation. Canonical encodings and a digest let the observer compare the approved proposal with the attempted action. The observer records the fields each path checks. The experimental adapters and any change to enforcement, persistence, or retry logic remain part of the declared configuration.

A repeated dispatch of the same intended effect retains its operation identifier. A distinct effect receives a new identifier. The observer evaluates each write under the authority applicable at its dispatch boundary and each reconciliation lookup under its read authority. The time of any change to a grant is recorded separately from the time of the original commitment. Provider deduplication can prevent a second effect even when a duplicate write request reaches the provider.

\subsection{Stress conditions and fault schedule}

\begin{figure}[htbp]
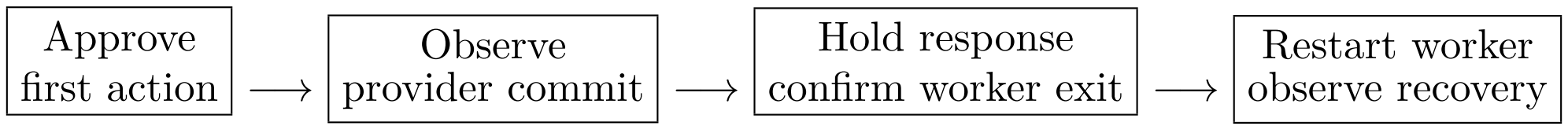

\centering
\small
\fbox{\shortstack{Approve\\first action}}
\(\longrightarrow\)
\fbox{\shortstack{Observe\\provider commit}}
\(\longrightarrow\)
\fbox{\shortstack{Hold response\\confirm worker exit}}
\(\longrightarrow\)
\fbox{\shortstack{Restart worker\\observe recovery}}
\caption{Post-commit fault sequence used in the paired pilot.}
\label{fig:timeline}
\end{figure}

The completed response-loss pilot used report delivery and paired the configurations under a normal E1 response, an E1 recipient change before dispatch, and post-commit response loss in E1 and E0. In the fault conditions the supervisor observed a durable provider commitment, held the response, terminated the designated worker, confirmed process exit, and restarted from persisted state. The observer checked the process and provider events for this order and recorded the task, assignment, approval and effect evidence before and after restart. The applicable grant remained in place in the fault conditions. Figure~\ref{fig:timeline} shows the observed sequence.

The proposed extension crosses approval revision, response loss, and provider mode as shown in Table~\ref{tab:cells}. In the combined condition the supervisor records the revision after the first commitment and before any possible recovery dispatch. Withdrawal and replacement with a separately approved action are distinct variants.

\paragraph{Provider facilities.}
E1 supports durable deduplication and lookup by the original operation identifier. Under E0 the application lacks both facilities. Both modes preserve an observer-only ledger. The observer distinguishes a receipt-backed recovery from a stop for manual reconciliation and checks for duplicate requests and effects in each outcome.

\subsection{Outcomes and attribution}

For each task, the grader links authorization decisions, attempts, commits, responses, and terminal states. It reports the following outcomes with explicit eligibility and denominators.

\begin{itemize}
    \item \textbf{Disallowed write request.} A write request reached the mock provider's ingress with no applicable authorization. The observer uses the request before provider deduplication and the independently recorded decision and assignment in force for that dispatch. An allowed reconciliation lookup and a product refusal recorded before dispatch have their own observations.
    \item \textbf{Unauthorized committed effect.} At least one committed change lacked applicable authorization at the declared boundary. Provider deduplication can yield a disallowed write request with zero new commits.
    \item \textbf{Duplicate effect.} At least two committed effects realize one logical operation. The grader checks ledger multiplicity by operation identifier as well as final state.
    \item \textbf{Correct completion.} All effects required by the declared task and policy occur with the expected multiplicity, and the final business state matches the manifest. An observation-only wait has a separate eligibility rule.
    \item \textbf{Recovery outcome.} Automatic recovery, confirmed safe stop requiring manual reconciliation, and observed failure are recorded separately. The grader uses an unknown outcome for metrics whose required evidence is missing.
\end{itemize}

When evidence is incomplete, a positively established violation remains reportable. Clean outcomes require the data specified for that metric. For an outcome among eligible trials, we report the unknown fraction and the range obtained by considering every unknown first negative and then positive. This keeps missing observations visible alongside the observed rate. End-to-end attribution uses the earliest observed divergence among approved proposal, attempted action, provider commitment, and recorded completion.

\subsection{Comparative decision criteria}
\label{sec:comparative-criteria}

For the completed pilots we report work acceptance, authorized completion and evidence, observed provider requests and effects, and the location of each exercised control in the product or application. Each pilot has one observation per condition and configuration. We describe paired differences in that fixture and keep deployed observations separate.

\subsection{Prospective deployment and held-out evaluation}

The proposed deployment comparison submits identical work and scope to a designated, pre-registered executor in each configuration. It observes each sender's dispatch and the deployed executor's receipt or claim separately. The grader links the original business task, intended and actual executor identities, authority, and result receipt. A read-only assignment control measures whether a deployed receiver with an assigned task makes any protected write requests. A separate authorized control measures successful writing through a supported product path. Reviewers then use equivalent task histories and comparable operator permissions to identify the responsible runner, the reason a task is waiting or complete, the relevant decision, and the result evidence. The grader checks those facts against independent records and reports evidence coverage. The deployed reference may use Agent Server or a host-side adapter, with the operator surface and version fixed in advance.

The proposed state-change controls cover approval from another task, reassignment after lease expiry, pre-dispatch revocation, and post-commit revocation or replacement. The supervisor will qualify event-time grading with a valid original commitment followed by a request under a revoked grant. The grader must flag that second request even when deduplication prevents a new effect. A second fixture will verify scoring of an approved replacement. An intent-matching control will check detection of duplicates when the application changes its operation identifier and distinguish two independent authorized actions with identical inputs.

\begin{table}[htbp]
\small
\centering
\begin{tabularx}{\textwidth}{@{}llX@{}}
\toprule
Approval & Execution & Diagnostic purpose \\
\midrule
Maintained & Normal response & Verify one intended effect and the expected final state. \\
Revised & Normal response & Evaluate a change of scope under ordinary execution. \\
Maintained & Post-commit failure & Verify recovery of an unacknowledged commitment. \\
Revised & Post-commit failure & Verify reconciliation of the earlier effect and the authority for any subsequent write. \\
\bottomrule
\end{tabularx}
\caption{Proposed stress matrix by eligible workload and provider mode.}
\label{tab:cells}
\end{table}

The primary prospective endpoint is the proportion of tasks with a disallowed write after a verified E1 fault and approval revision. We will report scheduled pairs, verified fault and revision exposure, missing observations, and correct completion alongside the paired rate difference $p_{\mathrm{AZOWL}}-p_{\mathrm{reference}}$. Eligibility requirements and thresholds for enforcement improvement and retained completion will be fixed before collection. The held-out study will freeze software, policy, replay rules, model configuration where applicable, fault schedules, fixtures, observation windows, and criteria for apparatus failures. Task failures and timeouts remain in the applicable denominators. Model-driven trials will record the requested model and runtime and the resolved version where verified. Repeated conditions from one fixture family will be treated as dependent.

The planned analysis reports matched differences by workload and provider mode and uses uncertainty intervals that preserve fixture dependence when the number of independent families permits. For each system and provider, the secondary interaction contrast for outcome $Y$ is
\begin{equation}
    \Delta_Y = (p_{11} - p_{10}) - (p_{01} - p_{00}),
    \label{eq:interaction}
\end{equation}
where $p_{ar}$ is the observed rate for approval revision $a$ and post-commit failure $r$, each coded as 0 or 1. The contrast will be compared between systems with cell-specific denominators and missing evidence. The earlier 1,080-run allocation is a design budget. The independent-family target depends on the desired precision and a prespecified meaningful difference.

\section{Results}
\label{sec:results}

The four completed studies comprise 14 matched case pairs and 28 runs. Table~\ref{tab:paired-results} reports the outcomes by study. Each pair compares two configurations on one prescribed condition. The studies share synthetic inputs and components, so their counts describe coverage of the tested conditions.

\begin{table}[!htbp]
\small
\centering
\begin{tabularx}{\textwidth}{@{}lrX@{}}
\toprule
Study & Pairs & Observed outcomes in the two configurations \\
\midrule
Human intake & 3 & Dialogue created no work task. Authorized submission produced one mock delivery after approval. An address change produced zero writes. \\
Agent-to-agent intake & 4 & Dialogue returned a reply with no work task. Unauthorized delegation created no task. Approved work delivered once. An address change produced zero writes. \\
Workflow reuse & 3 & The read returned a receipt. The mock provider recorded one dependent ticket after scope-matched approval. An assignee change produced zero writes. \\
Response-loss recovery & 4 & Both completed normal work and blocked changed scope. After response loss, E1 recovered the original receipt via one lookup and E0 stopped for manual reconciliation. Each fault trial retained one initial effect. \\
\bottomrule
\end{tabularx}
\caption{Paired laboratory studies on PC2. Each row counts one run per configuration for each listed condition. All protected writes targeted mock providers.}
\label{tab:paired-results}
\end{table}

\subsection{Conversation and delegated work}

The human-intake study sent the same report request through the discussion and explicit work entry points. Each discussion run recorded a reply with zero business tasks, work dispatch records, and mock inbox requests. Each authorized work submission created one addressable task and one dispatch request or intent. The mock inbox observed zero requests while approval was pending and one scope-matched delivery after approval.

The A2A study replayed the same request from the respective sender in each configuration. It added a handoff without work-creation permission, which yielded zero tasks and zero dispatches. The AZOWL arm exercised the original signed Master callback and recipient checks. The experiment routed a fixture response through the device interface. Approved work generated a product dispatch for the registered PC recipient with execution tools disabled. The LangGraph arm ran separate sender and receiver graphs and recorded the receiver's pickup from its application queue. The two arms used a common laboratory service to link each handoff identifier to the approval and mock delivery. As a supplemental access observation, removing work-creation permission in the AZOWL PC-owner fixture invalidated the previously selected recipient for dialogue. Dialogue remained available in the configured LangGraph role. These probes used different recipient-access models.

The delivery-address change occurred after the initial approval and before dispatch. Each path issued zero writes. The configured reference stopped at a checkpoint mismatch. The AZOWL component generated a fresh question with the changed scope, and the control service recorded it under the original task identifier. The experiments observed the new question and preserved the original approval record. Completion after a second approval remains untested.

\subsection{Governance reuse across two applications}

An isolated probe of AZOWL's workflow services bound the same versioned provider to an inventory workflow and a dependent ticket workflow. The product rejected the ticket step when its predecessor was absent or unverified. After verification of the fixture predecessor receipt, the original service emitted a generic dispatch request with the predecessor task identifier. The experiment captured that request at the substituted runtime transport.

\paragraph{Laboratory ticket adapter probe.}
The generic dispatch carried \path{operator_tools_enabled=false}. This part of the probe measured submission after prerequisite verification. A separate input exercised the typed entry point for the registered inventory-read adapter and its read-only binding. The input declared the experimental identifier \path{internal.lab.ticket.create.v1}. The product rejected it in preparation, before system binding lookup, with \path{SYSTEM_OPERATION_BINDING_SCOPE_CHANGED}. The dispatch function then rejected the same input with \path{WORKFLOW_SYSTEM_EXECUTION_CHANGED}, with zero additional runtime submissions. This is a validation result for a laboratory adapter at a specific read-only entry point.

The paired reuse study assembled a laboratory ticket input from the inventory receipt and proposed ticket fields. The AZOWL arm used the original compiled-plan gate and answer consumer through a laboratory bridge. After scope-matched approval, the laboratory connector sent one write to a mock ticket service, which stored the ticket in its local SQLite ledger. The service received zero writes when the assignee changed. The configured LangGraph app reused its graph factory, read node, and checkpoints. For tickets it attached its existing approval and journal subgraph and the mock provider connection. Table~\ref{tab:control-placement} attributes the controls by path.

\begin{table}[!htbp]
\small
\centering
\begin{tabularx}{\textwidth}{@{}>{\raggedright\arraybackslash}p{0.15\textwidth}>{\raggedright\arraybackslash}X>{\raggedright\arraybackslash}X@{}}
\toprule
Boundary & Original AZOWL code exercised & Laboratory or comparison-app contribution \\
\midrule
Work intake and routing & Isolated backend checked work-creation permission, recorded the task and recipient, and issued a read-only dispatch. & Synthetic callers, fixture reply, captured runtime transport. The LangGraph app provided its own intake, queue, and receiver. \\
Workflow binding & Product probe bound one versioned provider and checked the predecessor before generic submission. & Fixture predecessor receipt, substituted runtime transport, and experimental ticket adapter. \\
Assignment and approval & Both arms shared isolated original central task and approval services. The native scope gate and answer consumer ran as components. & The supervisor created the linked test assignment and supplied the gate inputs. The comparison app used a graph interrupt and scope checks. \\
Writes and recovery & Scope-gate decisions were observed at the connected pre-dispatch boundary. & Experimental tool adapters, mock provider ledger, fault injection, and an operation journal for each configured path. \\
\bottomrule
\end{tabularx}
\caption{Attribution of checks in the paired pilots and product workflow probe. The deployed cases are reported separately.}
\label{tab:control-placement}
\end{table}

\subsection{Recovery after response loss}

The recovery study began after task admission. In four fault exposures, the supervisor observed one provider commit and stopped the worker while the acknowledgement was withheld. A fresh worker resumed from persisted operation state. With lookup available (E1), each configuration recovered its original receipt through one query. With lookup unavailable (E0), each reached manual reconciliation. Across both fault conditions, the provider observed zero additional writes after restart. The AZOWL arm used the original scope gate with an experimental delivery journal. The reference combined graph checkpoints with its application operation journal.

\subsection{Observations from deployed AZOWL paths}

A read-only audit examined two completed portal conversations on 16 November 2026 (KST). Each had a native transport record and a lease for its reply. The audit found zero corresponding business tasks and zero links from the conversation to a business task. In a separate direct A2A probe on 17 November, a registered internal sender received one logical reply from the targeted team agent. Independent readback matched the request and reply identifiers. The inspected task indexes contained zero probe-related jobs.

The initial live inventory run on client~0.6.53 returned eight assets with a receipt linked to the assigned runner. On~0.6.54, the runner claimed a fresh copy, recorded a model capability error, and had no receipt at the observation cutoff. On~0.6.56, role and host checks passed after role synchronization, but execution stopped before model launch because the shared profile lacked the caller credential path. Client~0.6.57 resolved that path from the selected runner's enrollment.

On 19 November 2026 (KST), one fresh 0.6.57 run with the same provider revision yielded the end-to-end read trace in Table~\ref{tab:deployed-read-trace}.

\begin{table}[!htbp]
\small
\centering
\caption{Trace of the 0.6.57 PC2 inventory read (task alias \texttt{READ-01}).}
\label{tab:deployed-read-trace}
\vspace{4pt}
\begin{tabularx}{\textwidth}{@{}>{\raggedright\arraybackslash}p{0.21\textwidth}>{\raggedright\arraybackslash}X@{}}
\toprule
Stage & Verified evidence \\
\midrule
Submission & The central service accepted one workflow execution request. \\
Runner pickup & The PC2 pickup record links the queued task to its assigned runner. \\
Tool execution & One completed native inventory-read call carried a receipt reference. \\
Receipt and hashes & The central service verified the same receipt and the current lease at completion. Local and central result and evidence hashes matched. The read returned eight assets in one typed record. \\
\bottomrule
\end{tabularx}
\end{table}

On 4 November 2026, a report task waited for a scoped decision and finished on the same runner. Four downloaded artifacts matched the runner hashes. The business semantics review passed and the overall review flagged an evidence row count and a missing synthetic-data label.

On 12 November 2026, the deployed GRC audit verified idempotent sample ticket registration. An unresolved discrepancy blocked completion after approval, and completion persisted after remediation and renewed approval. On 13 November, the deployed GRC diagnostic service derived two follow-up proposals from an AZOWL runner's verified analysis. An operator-authenticated audit script registered one proposal as a sample ticket through the deployed GRC API and retrieved it by its returned ID.

\section{Limitations and Reproducibility}

The paired studies use synthetic cases with one run per configuration in each pair. The original AZOWL intake, workflow, and scope checks support claims at their tested boundaries. All protected writes in the paired pilots used laboratory adapters and mock providers. The live PC2 observations cover bounded reads on one host and sit outside the paired evaluations. The first 0.6.57 native evidence collection stopped with a missing-file error. A subsequent readback of the exact task artifacts completed the evidence join for the same run.

The deployed GRC cases cover sample tickets registered under operator authority. A read-only readiness check on 18 November 2026 found the deployed PC2 workflow provider ready with the registered system operation for inventory reads. Its permitted tools lacked a ticket creation operation. An end-to-end ticket trial with this runner would require a registered write tool that checks the task assignment and approved ticket input. Completion after renewed approval and approval withdrawal during acknowledgement loss remain open.

\paragraph{Reproducibility.}
The laboratory used Python~3.12.10, LangGraph~1.2.22, and SQLite checkpointer~3.1.3. The paired studies pin the central service at \texttt{00ad88ab} and released scope-gate source at \texttt{5f13f399}. The human-intake backend was \texttt{d4438f33}, the A2A backend was \texttt{284a7c5}, and the workflow registration probe used \texttt{6baecd59}. The completed live PC inventory cases used clients~0.6.53 and~0.6.57. The follow-up pins desktop source \texttt{75be8a28}, harness source \texttt{547babb6}, and provider revision~2. The internal research bundle records dependency locks, per-batch source hashes, the fixture and policy, task and approval traces, provider logs and ledgers, and independent verification. The GRC audits are retained separately with their live HTTP verification summaries. These artifact records have no public release accompanying this draft. The four local entry points are as follows.

\begin{quote}
\small\ttfamily
python -m azowl\_lab.intake\_comparison\\
python -m azowl\_lab.a2a\_comparison\\
python -m azowl\_lab.reuse\_comparison\\
python -m azowl\_lab.w01\_rehearsal --experimental-connector
\end{quote}

\section{Conclusion}

AZOWL AgentOps separates conversational exchanges from explicit work delegation for human and agent senders. Accepted work enters a common task lifecycle across registered execution environments. The default customer PC profile is read-only. A sender's authority to delegate, a runner's ownership under an assignment lease, and permission for a protected action are separate checks. The task identifier links those decisions to the records available for review.

Both pilot configurations kept dialogue free of business tasks, blocked unauthorized handoffs, and stopped mock writes after a change in approved scope. AZOWL used existing work-admission, routing, and prerequisite checks. The LangGraph reference supplied business task admission and dispatch in application code. The workflow probe established provider reuse and prerequisite enforcement within AZOWL. In the paired study, the ticket write exercised scope enforcement through a laboratory connector. A deployed PC2 follow-up verified a bounded inventory read under the repaired shared runtime. The deployment audit separately verified the operator-driven GRC ticket lifecycle on sample records. The observed results are scoped to those product boundaries and configured paths.

\section*{Use of Generative AI}

OpenAI Codex assisted with drafting and editing, parts of the experimental apparatus and verification code, and the synthetic sensitivity analysis. The named human authors are responsible for verifying the manuscript, methods, data, references, and conclusions before submission.

\appendix
\section{Additional qualification studies}

\paragraph{Apparatus and intake checks.}
The recorded apparatus qualification contained 54 automated checks covering persistence, the grader, concurrent sink attempts, and a central control restart. A scripted reference qualification covered 15 runs over three workloads and five conditions. Six runs terminated a worker after commit and before acknowledgement. An earlier batch contained three E0 outcomes with incomplete trace evidence. A separate reference replay covered four report-delivery conditions. Five isolated tests on backend \texttt{d4438f33} exercised read-only tenant chat, durable work intake, and server-selected discussion mode. These observations support the laboratory instruments and the tested intake paths.

\paragraph{Scope checks and earlier development runs.}
On runtime source \texttt{5f13f399}, 22 isolated tests exercised scoped compiled-plan approval and withholding engine launch. A separate direct-command category-gate probe accepted a changed target within the approved category. The probe issued no command. Two connected report checks observed one approved delivery and zero requests under a changed address. An earlier paired development batch covered eight runs across four conditions. Its AZOWL path reopened the approval gate after restart, with business recovery unobserved because the experimental connector was absent from that build. The primary recovery results in Table~\ref{tab:paired-results} use the later connector.

\paragraph{Continuity at individual boundaries.}
An isolated restart of the central control service preserved a pending question and its lease. On PC2, two isolated runs of the native question and answer-ingest functions reached \texttt{prompt\_ready}, including one that restarted while the question was pending. A separate three-run PC2 connector check observed normal delivery, an E1 receipt lookup after interruption, and an E0 handoff for manual reconciliation. In an artifact-storage trial, the original central service accepted an upload whose response was dropped. The artifact was recovered by reading it after restart, with one artifact record and one upload event.

\paragraph{Sensitivity exercise.}
A deterministic sensitivity exercise enumerated 200,000 synthetic paths across 20 assumption cells. As an illustration, when the original operation identifier was available in 90\% of already-committed E1 paths, lookup resolved that fraction under the assumed provider behavior. The counts describe the scenario grid and are reported separately from empirical trials.

\begingroup
\small
\bibliographystyle{plain}
\begingroup
\raggedright
\bibliography{references}

\begin{thebibliography}{10}

\bibitem{deng2026auditable}
Bangjie Deng, Xiaoxuan Lu, Wenjun Chen, Chengge Wen, Yulin Wang, Jiayue Meng,
  Yao Yu, Liying Liang, Weijia Han, Ziwei Hong, and Zidong Yu.
\newblock Auditable authorization for {LLM} agents: Empirical evidence across
  model, harness, and workflow layers.
\newblock \url{https://arxiv.org/abs/2611.19237v1}, 2026.
\newblock arXiv preprint, version 1.

\bibitem{langchain2026a2a}
{LangChain}.
\newblock {A2A} endpoint in {Agent Server}.
\newblock \url{https://docs.langchain.com/langsmith/server-a2a}.
\newblock Accessed 19 November 2026.

\bibitem{langgraph2026fault}
{LangChain}.
\newblock Fault tolerance.
\newblock
  \url{https://docs.langchain.com/oss/python/langgraph/fault-tolerance}.
\newblock Accessed 19 November 2026.

\bibitem{langchain2026platform}
{LangChain}.
\newblock {LangGraph} overview.
\newblock \url{https://docs.langchain.com/oss/python/langgraph/overview}.
\newblock Accessed 19 November 2026.

\bibitem{langchain2026server}
{LangChain}.
\newblock {LangSmith Agent Server}.
\newblock \url{https://docs.langchain.com/langsmith/agent-server}.
\newblock Accessed 19 November 2026.

\bibitem{langchain2026observability}
{LangChain}.
\newblock {LangSmith Observability}.
\newblock \url{https://docs.langchain.com/langsmith/observability}.
\newblock Accessed 19 November 2026.

\bibitem{langchain2026runs}
{LangChain}.
\newblock Runs.
\newblock \url{https://docs.langchain.com/langsmith/runs}.
\newblock Accessed 19 November 2026.

\bibitem{lee2015rifl}
Collin Lee, Seo~Jin Park, Ankita Kejriwal, Satoshi Matsushita, and John
  Ousterhout.
\newblock Implementing linearizability at large scale and low latency.
\newblock In {\em Proceedings of the 25th Symposium on Operating Systems
  Principles}, pages 71--86. ACM, 2015.
\newblock
  \href{https://web.stanford.edu/~ouster/cgi-bin/papers/rifl.pdf}{Author copy}.
  DOI \nolinkurl{10.1145/2815400.2815416}.

\bibitem{ly2015compliance}
Linh~Thao Ly, Fabrizio~Maria Maggi, Marco Montali, Stefanie Rinderle-Ma, and
  Wil M.~P. van~der Aalst.
\newblock Compliance monitoring in business processes: Functionalities,
  application, and tool-support.
\newblock {\em Information Systems}, 54:209--234, 2015.
\newblock \url{https://doi.org/10.1016/j.is.2015.02.007}.

\bibitem{mei2025aios}
Kai Mei, Xi~Zhu, Wujiang Xu, Mingyu Jin, Wenyue Hua, Zelong Li, Shuyuan Xu,
  Ruosong Ye, Yingqiang Ge, and Yongfeng Zhang.
\newblock {AIOS}: {LLM} agent operating system.
\newblock In {\em Conference on Language Modeling}, 2025.
\newblock
  \href{https://proceedings.colm.cc/paper_files/paper/2025/hash/fb575ab4d882a4c734641155a5f30911-Abstract-Conference.html}{COLM
  proceedings}.

\bibitem{palumbo2026forge}
Nils Palumbo, Sarthak Choudhary, Jihye Choi, Guy Amir, Prasad Chalasani, Isil
  Dillig, and Somesh Jha.
\newblock Formal policy enforcement for agentic systems.
\newblock \url{https://arxiv.org/abs/2602.16708v4}, 2026.
\newblock arXiv preprint, version 4.

\bibitem{sah2026undobench}
Dolly Sah, Tanmay Sah, Harshul Jain, and Tanya Sah.
\newblock {UndoBench}: Separating task competence from recovery capability in
  tool-using {AI} agents.
\newblock \url{https://arxiv.org/abs/2610.05622v1}, 2026.
\newblock arXiv preprint, version 1.

\bibitem{santos2026temporary}
Igor Santos-Grueiro.
\newblock Temporary authority, permanent effects: Commit-time authorization for
  {LLM} agents.
\newblock \url{https://arxiv.org/abs/2607.10487v2}, 2026.
\newblock arXiv preprint, version 2.

\bibitem{thomas1998tbac}
Roshan~K. Thomas and Ravi~S. Sandhu.
\newblock Task-based authorization controls ({TBAC}): A family of models for
  active and enterprise-oriented authorization management.
\newblock In T.~Y. Lin and Shelly Qian, editors, {\em Database Security XI:
  Status and Prospects}, pages 166--181. Springer, 1998.
\newblock Presented at the 1997 IFIP workshop.
  \url{https://doi.org/10.1007/978-0-387-35285-5_10}.

\bibitem{wang2026agentspec}
Haoyu Wang, Christopher~M. Poskitt, and Jun Sun.
\newblock {AgentSpec}: Customizable runtime enforcement for safe and reliable
  {LLM} agents.
\newblock In {\em Proceedings of the 2026 IEEE/ACM 48th International
  Conference on Software Engineering}, pages 2938--2950. ACM, 2026.
\newblock
  \href{https://cposkitt.github.io/files/publications/agentspec_llm_enforcement_icse26.pdf}{Author
  copy}. DOI \nolinkurl{10.1145/3744916.3764546}.

\bibitem{yao2025taubench}
Shunyu Yao, Noah Shinn, Pedram Razavi, and Karthik Narasimhan.
\newblock {\ensuremath{\tau}-bench}: A benchmark for tool-agent-user
  interaction in real-world domains.
\newblock In {\em International Conference on Learning Representations}, 2025.
\newblock
  \url{https://proceedings.iclr.cc/paper_files/paper/2025/hash/1b126cc38b8638e07bef37e7b2bb72bf-Abstract-Conference.html}.

\bibitem{zhang2025aflow}
Jiayi Zhang, Jinyu Xiang, Zhaoyang Yu, Fengwei Teng, Xiong-Hui Chen, Jiaqi
  Chen, Mingchen Zhuge, Xin Cheng, Sirui Hong, Jinlin Wang, Bingnan Zheng, Bang
  Liu, Yuyu Luo, and Chenglin Wu.
\newblock {AFlow}: Automating agentic workflow generation.
\newblock In {\em International Conference on Learning Representations}, 2025.
\newblock
  \url{https://proceedings.iclr.cc/paper_files/paper/2025/hash/5492ecbce4439401798dcd2c90be94cd-Abstract-Conference.html}.

\bibitem{zhu2026reconstruction}
Junchi Zhu, Zhenguang Liu, Shaojing Fan, Jianhai Chen, and Qinming He.
\newblock From approval to execution: Reconstruction-aware repair analysis for
  {LLM}-agent software.
\newblock \url{https://arxiv.org/abs/2609.26529v1}, 2026.
\newblock arXiv preprint, version 1.

\end{thebibliography}
\endgroup
\endgroup
\end{document}